\documentclass[a4paper,11pt]{article}
\pdfoutput=1 

\usepackage{jcappub} 

\usepackage[T1]{fontenc} 

\title{\boldmath Stable Gravastar model in Cylindrically Symmetric Space-time}

\author[1]{D. Bhattacharjee,}
\author[2]{P. K. Chattopadhyay,}
\author[3]{B. C. Paul.}

\affiliation[1,2]{IUCAA Centre for Astronomy Research and Development (ICARD), Department of Physics, Coochbehar Panchanan Barma University, Vivekananda Street, District: Coochbehar, \\ Pin: 736101, West Bengal, India}
\affiliation[3]{IUCAA Centre for Astronomy Research and Development (ICARD) and Department of
	Physics, University of North Bengal, West Bengal 734014, India.}

\emailAdd{debadriwork@gmail.com}
\emailAdd{ pkc$_{-}$76@rediffmail.com}
\emailAdd{bcpaul@associates.iucaa.in}

\abstract{We represent a class of new Gravastar solutions as obtained by Mazur and Mottola in a gravitational Bose-Einstein condensate (GBEC) in a cylindrical symmetric space-time. A stable gravastar with three distinct regions namely, (i) Interior de-Sitter space, (ii) Intermediate thin shell with a slice of finite length and (iii) exterior vacuum region. The interior region is characterised by positive energy density and negative pressure $(p=-\rho)$, which exerts a repulsive outward force at all points on the thin shell. The thin shell separating the interior and exterior is supposed to be consisting of ultra-relativistic stiff fluid having equation of state $p=\rho$, which satisfies the Zel'dovich's criteria \cite{Z,Z1}. This thin shell, which is considered as the critical surface for the quantum phase transition, replaces both the classical de-Sitter and Schwarzschild event horizons. The new solution is free from any singularities as well as any information paradox. The energy density, total energy, proper length, mass and entropy of the shell region are explored in this model and the gravastar model is stable and physically viable.}

\begin{document}
\maketitle
\flushbottom

\section{Introduction}\label{intro}
In the year 1915, Einstein introduced the theory of gravity known as the general relativity. The cast-iron prediction of general relativity is the existence of Black hole. Through the study of stellar evolution, we have come to know that black holes are the end state of gravitational collapse of a stellar body. In 1916, Karl Schwarzschild found the first intriguing solution of general relativity that would characterise a black hole. The radius of a Schwarzschild black hole is expressed as $R_s=\frac{2GM}{c^2}$. In the convention $G=c=1$, this relation becomes $R_s=2M$. Many researchers since then have studied the properties of black holes and expressed great insights. However, there persists two main problems with the black hole theories. Firstly, at the centre ($r=0$), a singularity exists where the general relativity breaks down and secondly, the boundary of a black hole termed as the event horizon is a peculiar region. Classically, the situation for a particle at horizon remains the same but under the semi-classical or quantum mechanical treatment the situation is very different and there is also the existence of information paradox. \\
To remove such ambiguities in the theory, Mazur and Mottola \cite{Mazur} proposed a multi-layered configuration for the end state of gravitational collapse. Initially, they proposed a five layer configuration which was effectively reduced to three layers by Visser and Wiltshire \cite{Visser}. Mazur and Mottola model \cite{Mazur} gave an alternative definition of black hole formalism, expressing the compact object as Gravitational Vacuum Star or Gravastar. The model contains three layers-(i) the interior region, (ii) intermediate thin shell and (iii) the exterior region. The interior region is considered as the vacuum De-Sitter space which removes any notion of sigularity and the exterior region is the vacuum space-time. The possibility of an event horizon has been abdicated by the introduction of a thin shell with very small thickness. The thin shell separates the interior and exterior regions. Thus the problems with the presence of event horizon have also been removed. The equation of state (henceforth EoS) for the interior region is $p=-\rho$, which is consistant with the positive cosmological constant. The vacuum exterior region is characterised by $p=\rho=0$. The thin shell is hypothesized to conatin ultra-relativistic stiff matter and in this region the velocity of light and sound are equal \cite{Mazur}. The EoS of the thin shell is $p=\rho$. \\  
Chapline \cite{Chapline} applied quantum mechanics in this context while considering the horizon as a critical surface of phase transition. However, this very concept was thought to be the end state of stellar collapse by Gliner \cite{Gliner}. Inside the gravastar, the quantum fluctuations dominate $T_{r}^{r}$ and $T_{t}^{t}$ components of the energy-momentum tensor. Thereby stretching the EoS to the extreme end allowed by causality. This induces an instability in the interior region that leads to the cold Gravitational Bose-Einstein Condensate (GBEC). \\
In a recent study by Ray et al. \cite{Ray}, it has been shown that the existence of pressure anisotropy and obeying some particular energy conditions in GR is a necessary viable condition for a gravastar to be a physically sustainable configuration. Although, DeBenedictis et al. \cite{Benedictis}, have successfully established a stable gravastar model having continuous pressure and density. Celine et al. \cite{Celine} explored the gravastar configuration with pressure anisotropy without the infinitesimal thin shell approximation. Also some works on gravastars have been proposed in modified gravity theories. In the paper, Shamir and Ahmed \cite{Fara}, they have worked out gravastar configurations in Gauss-Bonnet gravity theory. Bhatti et al. \cite{MB} studied the model of gravastar in $f(g)$ gravity. Another study by Ghosh et al. \cite{Ghosh} showed the modeling of gravastars in $f(T,T)$ gravity theory.  \\
From the observational perspective, gravastars are getting much attention lately. The successful detection of gravitational waves in $2015$\cite{Ab}, originating from the merging of black hole binaries, do not comprehend the puzzles existing in case of a black hole. Therefore, an alternative approach is necessary. Regarding this backdrop, gravastar hypothesis can provide a soothing theory for the  end state of gravitational collapse. Even though gravastars have not been observed till now, it is crucial to study their properties to find out the missing piece of information to comprehend the understanding of black holes. Kubo and Sakai \cite{Kubo} in their study explored gravastars for an obeservational point of view. After comparing the maximal luminosity of a gravastar and a black hole of same mass, they \cite{Kubo} discovered that gravastar luminosity is greater. This is a significant result for the relavance of gravastars. Again, it is hypothesized that the $M87$ black hole image does not contain any event horizon and it can be considered a gravastar as per the existing concerned theory \cite{M87}. Cardoso et al. \cite{Cardoso} suggested that the ringdown signals of $GW150914$ detected by LIGO interfermetric studies are supposed to be generated by horizon-less compact objects. Thus, gravastars become much more significant in this context also. Quest for gravastars are still going on and interesting revelation are bound to happen. \\   
In this paper, we have adapted the Mazur-Mottola model in the most general cylindrically symmteric space-time form \cite{Bronnikov}. Our main objective for choosing the most general cylindrically symmetric space-time was to go beyond the conventional spherical concepts and find the relavant results regarding our model. We have constructed and solved the Einstein's field equations and imposing the conservation of energy on the solution we have achieved the coherent aforementioned results for the three layered model of the gravastar. From the junction conditions \cite{DI}-\cite{Israel1}, we have found out a comprehensive result for the mass of the shell of gravastar. We have also found that the entropy in our approach is much less than that of a classical black hole. Therefore, we may conclude that the model of the gravastar presented here may be considered as a viable alternative manifestation of black hole formalism. \\       
The paper is organised in distinct sections and subsections. In section (\ref{mathform}) we have considered the static cylindrically symmetric line element \cite{Bronnikov} and solved the Einstein field equations. Thereafter, we have imposed the gravastar structural conditions on the solution of field equations. By doing so, we have found out the behaviour of the solutions in the interior, thin shell and exterior regions respectively. Section (\ref{JC}) describes the junction condition, the energy density at the shell surface and surface pressure to obtain the mass of the gravastar. Section (\ref{IF}), depicts the total mass of the gravastar along with the fundamental characteristics of gravastar, namely, the proper length, energy and the entropy contained within the shell. In the section (\ref{eos}), we have shown the variation of EoS for thin shell $(p=\rho)$ with the shell thickness. Finally, in section (\ref{C}) we have discussed the main findings of our model. 
\section{Mathematical Formalism}\label{mathform}
\begin{figure}[tbp]
	\centering 
	\includegraphics[width=.45\textwidth]{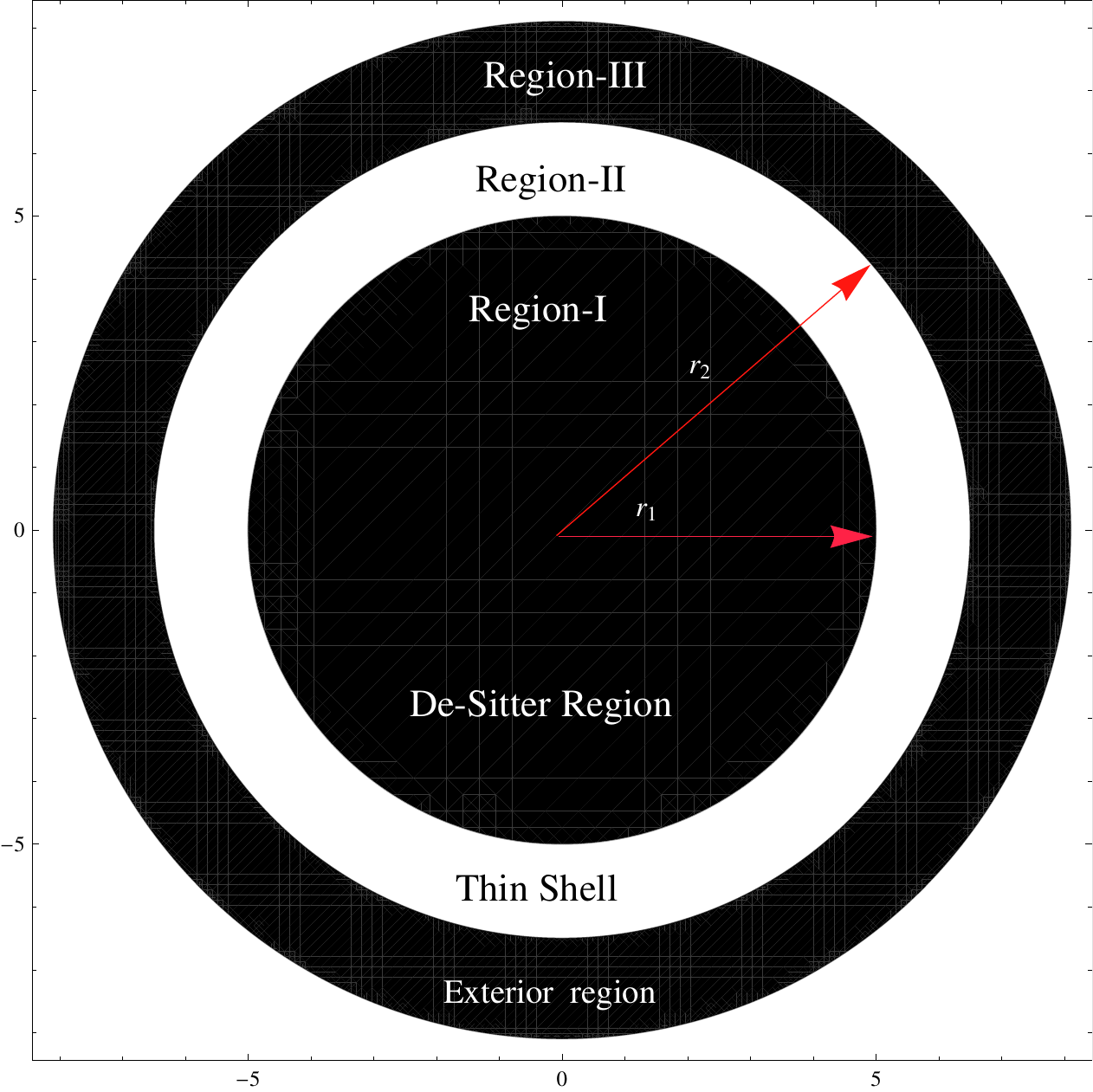}
	\caption{\label{fig1} Schematic diagram of Gravastar }
\end{figure}

We consider the most general static, cylindrically symmetric line element proposed by Bronnikov et al. \cite{Bronnikov} of the form as given below: 
\begin{center}
	\begin{equation}
		ds^2=e^{2\gamma} dt^2-e^{2\alpha} dr^2-e^{2\mu} dz^2-e^{2\beta} d\phi^2,\label{eq1}
	\end{equation}
\end{center}
here, $\gamma, \alpha, \mu, \beta$ are functions of $r$.
The Stress-Energy tensor for perfect isotropic fluid approximation takes the form:
\begin{center}
	\begin{equation}
		{T}^{\mu}_{\nu}=diag (\rho,-p,-p,-p).\label{eq2}
	\end{equation}
\end{center}
The Einstein field equation connecting geometry and matter segments is given as: 
\begin{center}
	\begin{equation}
		{R}^{\mu}_{\nu}-\frac{1}{2}{g}^{\mu}_{\nu}R=-8\pi {T}^{\mu}_{\nu}, \label{eq3}    
	\end{equation}
\end{center}
where, $R_{\nu}^{\mu}= $~Ricci tensor and $R=$~Ricci Scalar.
We have considered the system of units $G=c=1$. 
We have also considered a harmonic radial co-ordinate where $\alpha$ is represented in terms of the other three functions $\gamma, \mu, \beta$ as given below \cite{Bronnikov}:
\begin{center}
	\begin{equation}
		\alpha=\gamma+\beta+\mu,\label{eq4}
	\end{equation}
\end{center}
so that, the Einstein's field equations are tractable. 
Using Eqs.~\eqref{eq1} and \eqref{eq2}, the Einstein field equation given in Eq.~\eqref{eq3} reduces to the set of equations \cite{Bronnikov} given below,
\begin{equation}
	\beta''+\mu''-\mu'\beta'-\mu'\gamma'-\beta'\gamma'=-8\pi\rho e^{2\alpha},\label{eq5}
\end{equation}

\begin{equation}
	\mu'\beta'+\mu'\gamma'+\beta'\gamma'=8\pi pe^{2\alpha},\label{eq6}
\end{equation}
\begin{equation}
	\mu''+\gamma''-\mu'\beta'-\mu'\gamma'-\beta'\gamma'=8\pi pe^{2\alpha},\label{eq7}
\end{equation}
\begin{equation}
	\gamma''+\beta''-\mu'\beta'-\mu'\gamma'-\beta'\gamma'=8\pi pe^{2\alpha},\label{eq8}
\end{equation}
where prime ($'$) represents derivative with respect to $r$ and where ${T}^{\mu}_{\nu}$ is the stress-energy tensor for the comoving reference frame in case of perfect fluid approximation. The four velocity $u^u$ is given as, $u^{\nu}=(e^{-\gamma},0,0,0)$. Therefore, the stress-energy tensor, ${T}^{\mu}_{\nu}$ takes the form, 
\begin{equation}
	{T}^{\mu}_{\nu}=(p+\rho)u^{\mu}u_{\nu}-{\delta}^{\mu}_{\nu}p=diag(\rho,-p,-p,-p) \label{eq9}
\end{equation}
Conservation of the stress-energy tensor $(\nabla_{\nu} {T}^{\mu}_{\nu}=0)$ can be written as,    
\begin{equation} 
	p^{\prime}+(p+\rho)\gamma^{\prime}=0. \label{eq10}
\end{equation}
Now, in Eq.~\eqref{eq1}, we have four unknown variables to be determined but we have only three relations among them. In order to have a closed system of equations, we have to use three EoS for the three regions of gravastar as follows:
\begin{itemize}
	\item Interior (Region-I): \hspace{1cm} $0\le r < r_1$ ,\hspace{0.5cm} $p=-\rho$
	
	\item Shell (Region-II): \hspace{1cm}$r_1< r<r_2$,\hspace{0.8cm} $p=\rho$ 
	
	\item Exterior (Region-III): \hspace{1cm}$r_2< r$,\hspace{0.9cm} $p=\rho=0$ 
\end{itemize}
All the metric coefficients need to be continuous at the interfaces of a gravastar at $r=r_1$ and $r=r_2$ as shown in Fig.~\ref{fig1}. Here we need to require that the metric coefficients $\alpha,\gamma,\mu,$ and $\beta$ to be continuous, but the first derivatives of $\alpha,\gamma,\mu,\beta,$ and $p$ must be discontinuous as evident from Eqs.~\eqref{eq5}, \eqref{eq6} and \eqref{eq10}.\\
\begin{itemize}
	\item {\bf Interior Region:} We construct the solution for the interior de-Sitter space (Region-I) with the form of equation of state $p=-\rho=constant$. To eliminate the singularity at $r\rightarrow0$, the interior region is considered as a vacuum de-Sitter space. We have established the analytical solution for the interior region. A linear combination of Eqs.~\eqref{eq5} to \eqref{eq8} shows,
	\begin{equation}
		\gamma''=-8\pi\rho e^{2\alpha}, \label{eq10a} 
	\end{equation}
	\begin{equation}
		\beta''=-8\pi\rho e^{2\alpha}. \label{eq10b}
	\end{equation}
	Eqs.~\eqref{eq10a} and \eqref{eq10b} yields the form of $\beta$ as given below, 
	\begin{equation}
		\beta=\gamma+a_1r+\ln{r_0}, \label{eq11} 
	\end{equation}
where, $a_1$ and $r_0$ are integrating constants.
	Solving Eqs.~\eqref{eq7} and \eqref{eq8}, and using the value of $\beta$ from Eq.~\eqref{eq11}, we have determined the value of $\mu$ as,
	\begin{equation}
		\mu=\gamma+(a_1+b_1)r, \label{eq12} 
	\end{equation}
where $b_1$ is also an integrating constant,
	Finally from Eq.~\eqref{eq4} we get, 
	\begin{equation}
		\alpha=3\gamma+ar+\ln{r_0}, \label{eq13}
	\end{equation}
	where, 
	\begin{equation}
		a=2a_1+b_1=constant. \label{eq14}
	\end{equation}
	The form of $\alpha,\beta$ and $\mu$ are similar to that of \cite{Bronnikov}. Using Eqs.~\eqref{eq11} and \eqref{eq12} in Eq.~\eqref{eq6} with the regular symmetry axis condition $a_1+b_1=0$, we get:
	\begin{eqnarray}
		a=\frac{4}{24\pi pr_{0}^{2}}, \\ \label{eq14a}
		e^{2\gamma} = \frac{a}{(1+e^{2ar})^{2/3}}, \label{eq15} \\
		e^{2\alpha} = \frac{a^3r_0^2e^{2ar}}{(1+e^{2ar})^2}. \label{eq16}
	\end{eqnarray}
	In the above formulation, $a_1,b_1,a~(=a_1+b_1)$ and $r_0$ are arbitrary constants. Though $r_0$ is a constant, it represents length scale. From Eqs.~\ref{eq15} and \ref{eq16} we have found that both $g_{tt}$ and $g_{rr}$ are finite at $r=0$, which eliminates the possibility of singularity. Therefore, for a classical black hole, the issue of singularity at the center $(r=0)$ my be ruled out. \\ The active gravitational mass contained in the de-Sitter space (region-I) of radius $r=r_1$, can be written as:
	\begin{equation}
		M_{active}=\int_{0}^{r_{1}} 4\pi r^2 \rho dr=\frac{4}{3}\pi r_{1}^{3}\rho_{c}. \label{16a}
	\end{equation}
where $\rho_{c}=$ interior density, which is assumed to be equal to the critical density of the universe $(\rho=\rho_{c}\approx10^{-29} gm/cm^3)$ as per Mazur and Mottola hypothesis of gravastar \cite{Mazur}. Considering the radius of interior De-Sitter space of gravastar as $9, 10$ and $11$ km and $\rho_{c}=10^{-29}gm/cm^3$ , we have evaluated the mass of the interior region of gravastar as given below: 
\begin{eqnarray}
	M_{9}= 1.53\times10^{-44} M_{\odot} \nonumber \\
	M_{10}= 2.10\times10^{-44} M_{\odot} \nonumber \\
	M_{11}= 2.80\times10^{-44} M_{\odot}. \label{eq16b}
\end{eqnarray}
where $M_{\odot}$ is the solar mass. 
	\item {\bf Thin Shell:} The only non-vacuum region is Region-II, the thin shell of ultra-relativistic fluid. The equation of state of this region is $p=\rho$. In this formulation we have adapted the same approach as discussed in \cite{Bronnikov} and the following expressions have been established: 
	\begin{eqnarray}
		\beta=a_1r+\ln{r_0}, \label{eq17} \\
		\mu=(a_1+b_1)r, \label{eq18}\\
		\alpha=\gamma+ar+\ln{r_0}, \label{eq19}
	\end{eqnarray}
	where, $a$ is given in Eq.~\eqref{eq14}. 
	Using Eq.~\eqref{eq6}, we have determined the values of $\gamma$ and $\alpha$ for regular symmetry axis \cite{Bronnikov} as, 
	\begin{eqnarray}
		\gamma=\frac{4\pi p}{a^2}e^{2ar}, \label{eq21} \\
		\alpha=\frac{4\pi p}{a^2}e^{2ar}+ar+\ln{r_0}. \label{eq22}
	\end{eqnarray}   
	Using Eqs.~\eqref{eq21} and \eqref{eq22} in Eq.~\eqref{eq10}, we get the equation of state in the form, 
	\begin{equation}
		p=\rho=\frac{a^2}{8\pi}e^{-2ar}, \label{eq23}
	\end{equation}
	In the thin shell limit \cite{BCP}, any parameter that depends on r is very small as $r\rightarrow0$. Therefore, from Eqs.~\eqref{eq21} and \eqref{eq22}, we assert that, in the limit $r\rightarrow0$ , $g_{tt}$ and $g_{rr}$ are non-vanishing constants. The non-vanishing property proves that event horizon is absent in this approach which indicates that gravastar may be a viable alternative manifestation of black hole. \\
	\item {\bf Exterior Region:} The exterior region is the static vacuum region with the equation of state $\rho=p=0$. The vacuum solution indicates, $R_{\nu}^{\mu}=0$ which implies, 
	\begin{equation}
		\beta''=\gamma''=\mu''=0. \label{eq24}
	\end{equation}
	Therefore, 
	\begin{eqnarray}
		\beta=C_3r+C_3^0, \label{eq25} \\
		\gamma=C_4r+C_4^0, \label{eq26} \\
		\mu=C_5r+C_5^0. \label{eq27}
	\end{eqnarray}
	Then from Eq.~(\ref{eq4}) we get, 
	\begin{equation}
		\alpha=C_6r+C_6^0, \label{eq28}
	\end{equation}
	where, $C_3,C_4,C_5,C_6,C_3^0,C_4^0,C_5^0,C_6^0$ are the arbitrary integration constants and in Eq.~\eqref{eq28}, $C_6=C_3+C_4+C_5$ and $C_6^0=C_3^0+C_4^0+C_5^0$. With proper coordinate transformation, we can set, $C_3^0,C_4^0,C_5^0,C_6^0$ equal to zero. From the above calculations, the line element in exterior region becomes, 
	\begin{equation}
		ds^2=e^{2C_4r} dt^2-e^{2C_6r} dr^2-e^{2C_5r} dz^2-e^{2C_3r} d\phi^2. \label{eq29}
	\end{equation}
	To show that the exterior spacetime is flat, we have calculated the Kretschmann Scalar $R_k$. The general form of the Kretschmann Scalar $R_k$ is given in \cite{Henry}, 
	\begin{equation}
		R_k=R^{pqrs}R_{pqrs}, \label{eq29a}
	\end{equation}
where $R_{pqrs}$ is the Riemann curvature tensor. Applying Eq.~\eqref{eq29a} in Eq.~\eqref{eq29}, we compute $R_k$ as,  
	\begin{eqnarray}
		R_k=\frac{1}{e^{4C_6r}}(4(C_3^4+C_4^4+C_5^2(C_5-C_6)^2-2C_3^3C_6+\\ \nonumber
		C_4^2(C_5^2+C_6^2)+C_3^2(C_4^2+C_5^2+C_6^2))), \label{eq30}
	\end{eqnarray}
	From Eq.~\eqref{eq30}, it is evident that due to the presence of the term $(\frac{1}{r})$ on the right hand side. the value of $R_k\rightarrow0$ as $r\rightarrow\infty$, showing that the spacetime is vacuum, i.e. the region is flat. 
\end{itemize}

\section{Junction condition}\label{JC} Gravastars have a three layered configuration, i.e., interior region, exterior region and the thin shell. The thin shell connects the interior geometry to the exterior one. We have computed the junction condition by equating the interior solution to the exterior one at $r=R$ and develope some useful relations which permits the interior and exterior geometries to match smoothly. It is to be noted here, although the metric co-efficients are continuous at the junction of the configuration, their derivatives are not necessarily continuous. Using the conditions given by Darmois-Israel \cite{DI}-\cite{BCP}, we have computed the expressions of surface stresses at the interface of the junction. Using Lanczos equation \cite{Lanczos}-\cite{Musgrave} the expression for the intrinsic surface energy tensor ${S}_{ij}$ is expressed as, 
\begin{equation}
	{S}^{i}_{j}=-\frac{1}{8\pi} ({K}^{i}_{j}-{\delta}^{i}_{j}{K}^{k}_{k}), \label{eq31}
\end{equation}
here, ${K}_{ij}={K}^{+}_{ij}-{K}^{-}_{ij}$ ,where $(+)$ and $(-)$ sign indicate the value of $K_{ij}$ at the exterior and interior interfaces respectively. Now the form of second fundamental ${K}_{ij}^{\pm}$ is given below:
\begin{equation}
	{K}_{ij}^{\pm}=-{\eta}_{ij}^{\pm} (\frac{\partial^2 X_{\nu}}{\partial \zeta^{i} \partial \zeta^{j}}+{\Gamma}_{\alpha\beta}^{\nu} \frac{\partial X_{\alpha}}{\partial \zeta^{i}} \frac{\partial X_{\beta}}{\partial \zeta^{j}} ). \label{eq32}
\end{equation}
The double sided normal on the surface is defined as, 
\begin{equation}
	\eta_{ij}^{\pm}={\pm} ({g}^{\alpha\beta} \frac{\partial f}{\partial x\alpha} \frac{\partial f}{\partial x\beta})^{1/2}, \label{eq33}
\end{equation}
with $\eta^{\nu}\eta_{\nu}=1$. Following Lanczos equation \cite{Lanczos}-\cite{Musgrave}, the surface stress-energy tensor at the boundary of the interface is defined as, ${S}_{ij}=diag(\varrho, -\vartheta, -\vartheta, -\vartheta )$, where $\varrho$ and $\vartheta$ are the surface energy density and the surface pressure respectively. The form of parameters $\varrho$ and $\vartheta$ are evaluated as, 
\begin{eqnarray}
	\varrho=-\frac{1}{4\pi R}{(\sqrt{f(r)})}^{+}_{-}~, \label{eq34} \\
	\vartheta=-\frac{\varrho}{2}+\frac{1}{16\pi} (\frac{f'(r)}{\sqrt{f(r)}})^{+}_{-}, \label{eq35}
\end{eqnarray}
where, $f(r)^{+}_{-}$ represents the $g_{tt}$ components of the exterior $(e^{2C_4R})$ and interior $(\frac{a}{(1+e^{2ar})^{2/3}})$ regions respectively. Using Eqs.~\eqref{eq15} and \eqref{eq29}, Eqs.~\eqref{eq34} and \eqref{eq35} reduces to, 
\begin{equation}
	\varrho=-\frac{1}{4\pi R}(e^{C_4R}-\frac{\sqrt{a}}{(1+e^{2aR})^\frac{1}{3}}) , \label{eq36} \\
\end{equation}
\begin{eqnarray}
	\vartheta=\frac{1}{8\pi R}(e^{C_4R}-\frac{\sqrt{a}}{(1+e^{2aR})^\frac{1}{3}}) \\ \nonumber+\frac{1}{16\pi} (\frac{2C_4e^{C_4R}}{e^{C_4R}}+\frac{4a^{\frac{3}{2}}e^{2aR}}{3(1+e^{2aR})^\frac{4}{3}}).\label{eq37}
\end{eqnarray}
Now using Eq.~\eqref{eq36}, one can determine the mass of the gravastar as, 
\begin{equation}
	M_{shell}=4\pi R^2\varrho=R( \frac{\sqrt{a}}{(1+e^{2aR})^\frac{1}{3}}-e^{C_4R}). \label{eq38}
\end{equation}  
\section{Important features of gravastar in this model} \label{IF} In this section we have analysed some basic characteristics of gravastars in cylindrically symmetric spacetime.
\subsection{Mass of gravastar} 
\begin{figure}[tbp]
	\centering 
	\includegraphics[width=.45\textwidth]{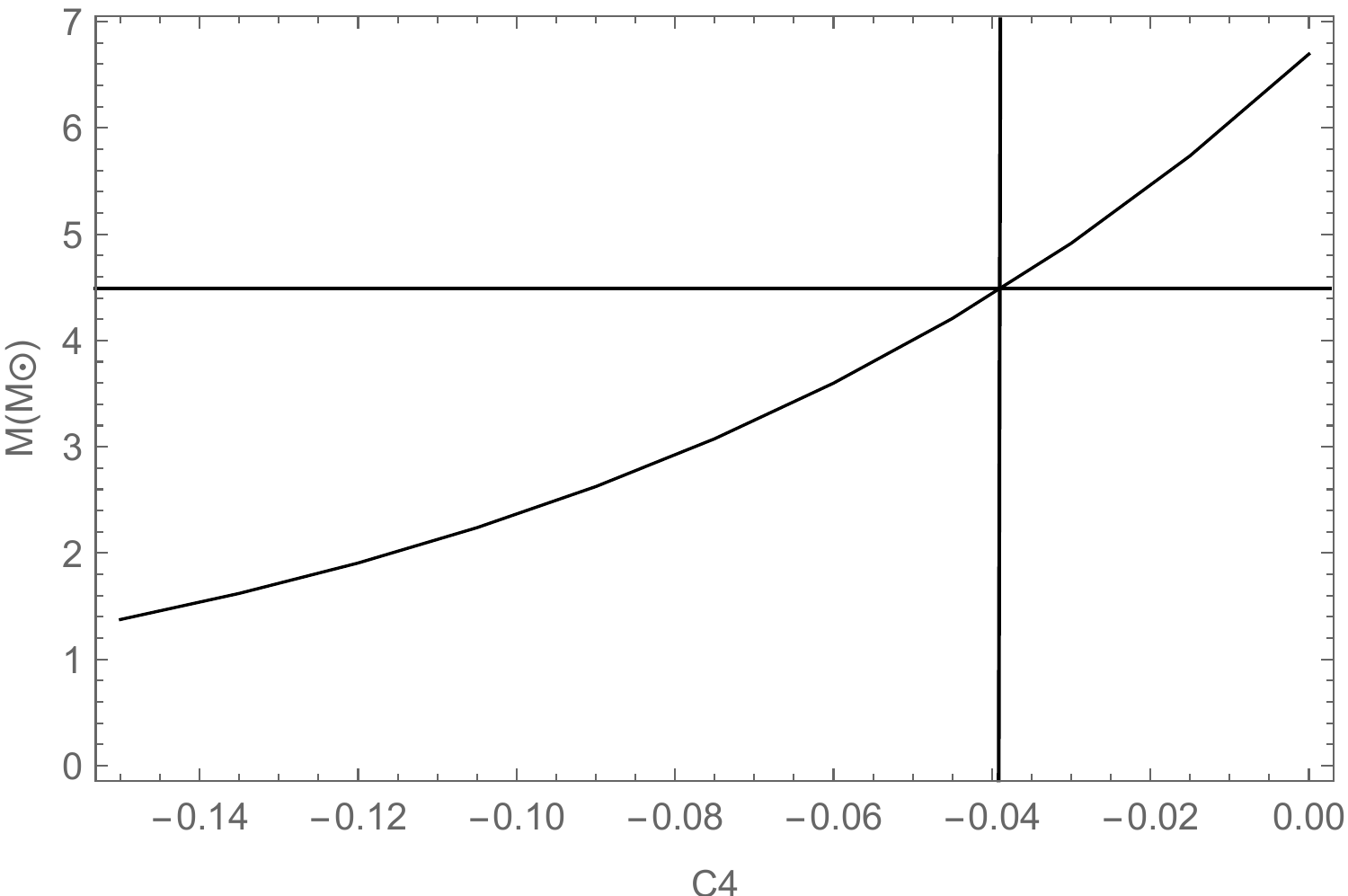}
	\caption{\label{fig2c}Variation of mass of the gravastar with $C_4$ with parametric choice of constant $a=0.0006$ and $R=9.07km$. }
\end{figure}

\begin{figure}[tbp]
	\centering 
	\includegraphics[width=.45\textwidth]{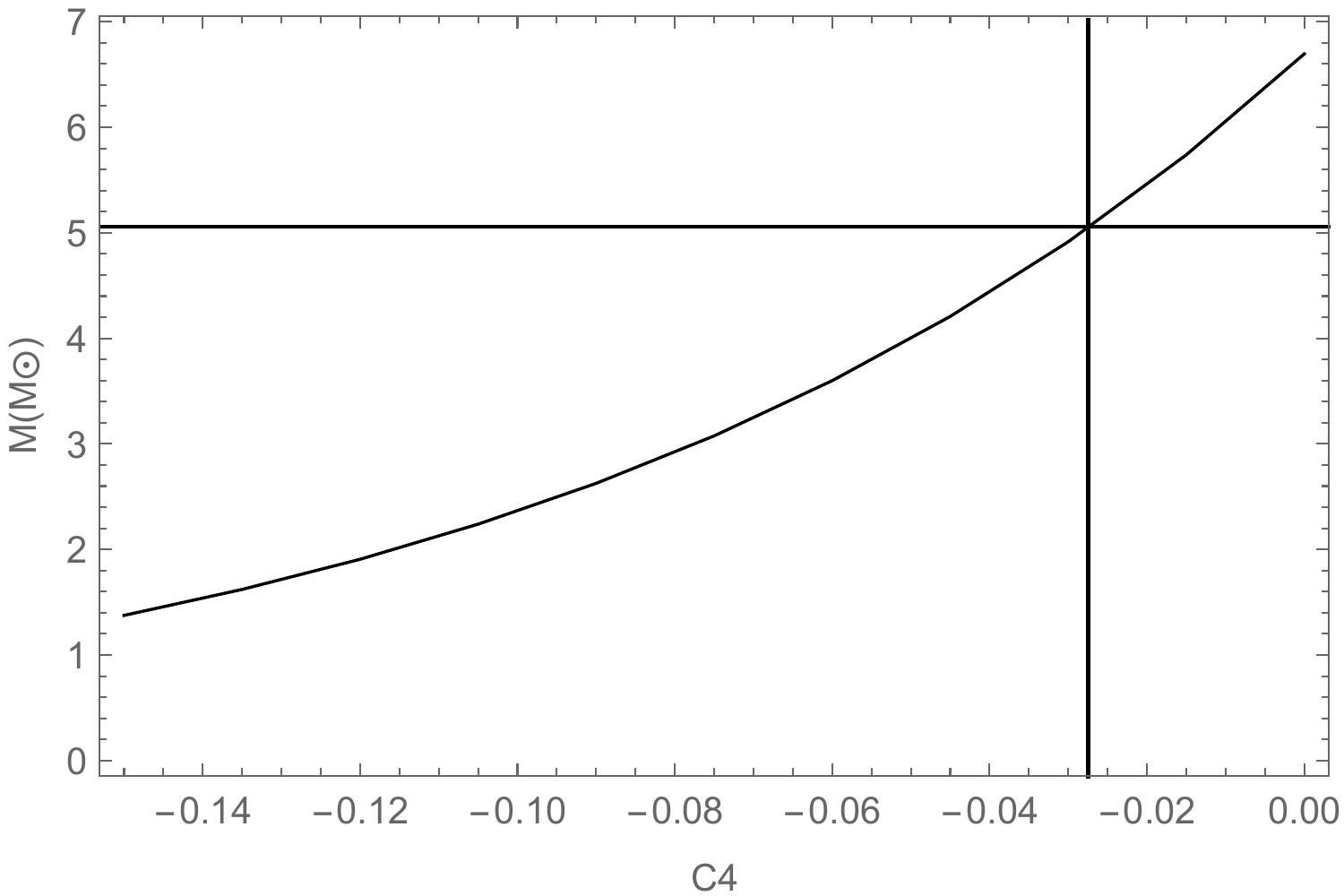}
	\caption{\label{fig2b}Variation of mass of the gravastar with $C_4$ with parametric choice of constant $a=0.0006$ and $R=10.07km$. }
	
\end{figure}

\begin{figure}[tbp]
		\centering 
		\includegraphics[width=.45\textwidth]{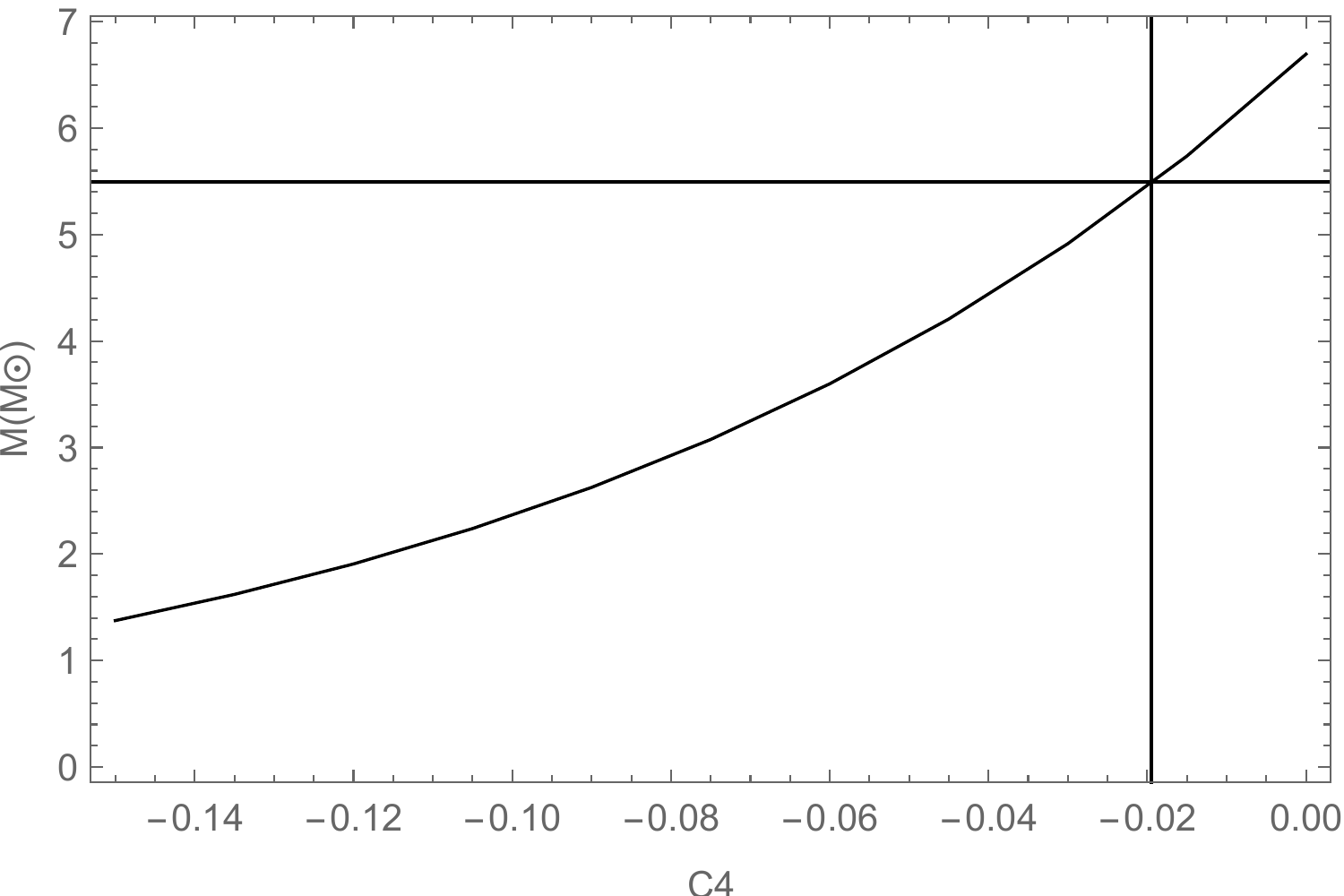}
		\caption{\label{fig2a}Variation of mass of the gravastar with $C_4$ with parametric choice of constant $a=0.0006$ and $R=11.07km$. }
	\end{figure}

In this section we have determined the possible mass contained within the thin shell of gravastar using Eq.~\eqref{eq38} and are shown in Figs.~(\ref{fig2c}-\ref{fig2a}). As $R>0$ and $a>0$, from Eq.~\eqref{eq36}, one may conclude that for any combination of $R$ and $a$,
$0<\frac{\sqrt{a}}{(1+e^{2aR})^\frac{1}{3}}<1$. Again, from Eq.~\eqref{eq38} as $M_{shell}>0$, we can write, 
\begin{equation}
	\frac{\sqrt{a}}{(1+e^{2aR})^\frac{1}{3}} > e^{C_4R}, \label{eq39}
\end{equation}
which implies that $C_4=-Ve$ always. We have considered the range of values of $C_4$ and plotted the dependence of mass of gravastar on $C_4$. From Figs.~\ref{fig2c}-\ref{fig2a}, some interesting results are noted. It is evident from Fig.~\ref{fig2c} that the mass-radius relation of gravastar follows the relation $R<2M$ when the value of $C_4$ lies in the range of $-0.039<C_4<0$. On the other hand $R>2M$ when $C_4<0.039$. Here $M$ represents the mass of gravastar within the radius $R$. Intersecting point indicates $R=2M$. From Figs.~\ref{fig2b} and \ref{fig2a}, it is also noted that the window of $C_4$ decreases as $R$ increases and above a certain value of $R$ it is not possible to get any value of $C_4$ so that we do not get any value of $M$. Interestingly it is evident that whether $R\le2M$ (condition for black holes) or $R>2M$, the concept of event horizon is absent. From Figs.~\ref{fig2c}-\ref{fig2a}, it is also noted that mass of gravastar decreases with the decrease of the value of $C_4$. It is evident from Figs.~\ref{fig2c}-\ref{fig2a} that the mass of the gravastar shell is much more than that of mass of region-I as given is Eq.~\ref{eq16b} for $r_1=9, 10$ and $11$ km respectively. So, it may be concluded that the mass of the gravastar means the mass of the thin shell as we can neglect the mass of region-I with respect to the mass of the thin shell. 
  
\subsection{Proper Length of the Shell}
The interior and exterior regions of the gravastar are separated by the thin shell. The EoS of the ultra-relativistic fluid within the shell, $p=\rho$, was first introduced by Zel'dovich \cite{Z,Z1} in the context of cosmology. Within the shell the velocity of sound and velocity of light are equal. The thickness of the shell is given as \cite{Mazur},

\begin{equation}
	\ell = \int_{r_1}^{r_2} e^\alpha dr, \label{eq40}
\end{equation}
Using Eqs.~\eqref{eq22} and \eqref{eq23}, 
\begin{eqnarray}
	\ell&=&r_0\int_{r_1}^{r_2} e^{(\frac{1}{2}+ar)} dr, \label{eq41} \\
		\ell/r_0&=&\frac{\sqrt{e}}{a}(e^{ar_2}-e^{ar_1}). \label{eq42} 
\end{eqnarray}

\begin{figure}[tbp]
	\centering 
	\includegraphics[width=.45\textwidth]{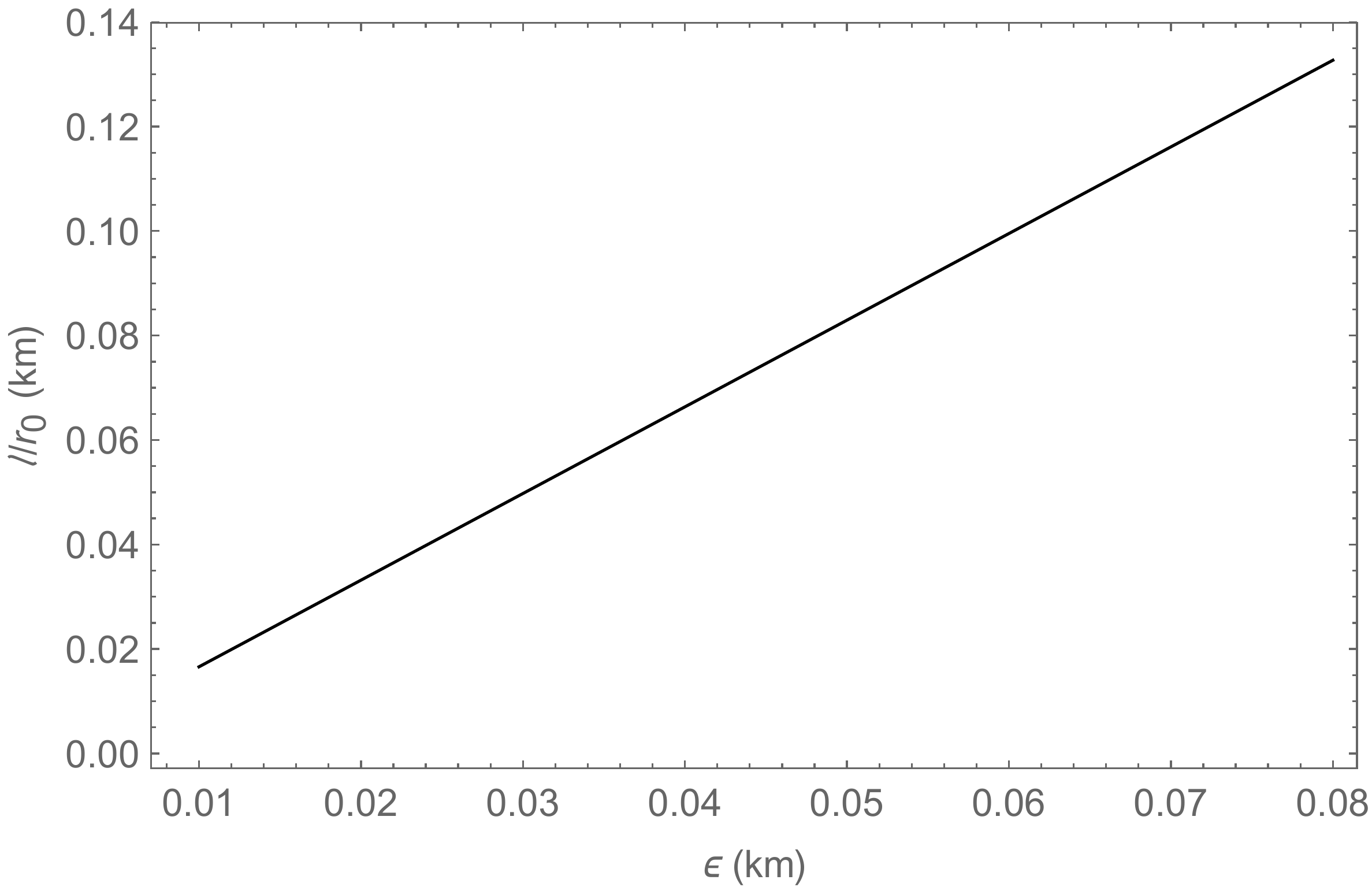}
	\caption{\label{fig3}Variation of Normalised Proper Length of the shell $(\ell/r_0)$ with Shell thickness $\epsilon=(r_2-r_1)$ for parametric choice of constant $a=0.0006$. }
\end{figure}
In Fig.~\ref{fig3}, we have shown the variation of normalized proper length $(\ell/r_0)$ with thickness $\epsilon=(r_2-r_1)$ of the thin-shell. It is evident that $\ell$ increases with the increase of thickness $\epsilon$ of the shell for any parametric choice of $a$. Similiar conclusions have also been expressed in  \cite{MB} and \cite{Yusaf}. 
\subsection{Energy of the Shell}
The EoS of the thin shell (Region-II) $p=\rho$ implies the stiff fluid approximation where the velocity of light is equal to the velocity of sound. The energy within the shell is given by: 
\begin{equation}
	E = 4\pi\int_{r_1}^{r_2} \rho r^2 dr, \label{eq42} 
\end{equation}
using Eq.~\eqref{eq23} we get, 
\begin{eqnarray}
	E&=&\int_{r_1}^{r_2} \frac{a^2}{2} r^2e^{-2ar} dr, \label{eq43} \\
	&=&\frac{e^{-2ar_1}}{4a^3}(1+2ar_1+2a^2r_{1}^{2})-\frac{e^{-2ar_2}}{4a^3}(1+2ar_2+2a^2r_{2}^{2})
\end{eqnarray}

\begin{figure}[tbp]
	\centering 
	\includegraphics[width=.45\textwidth]{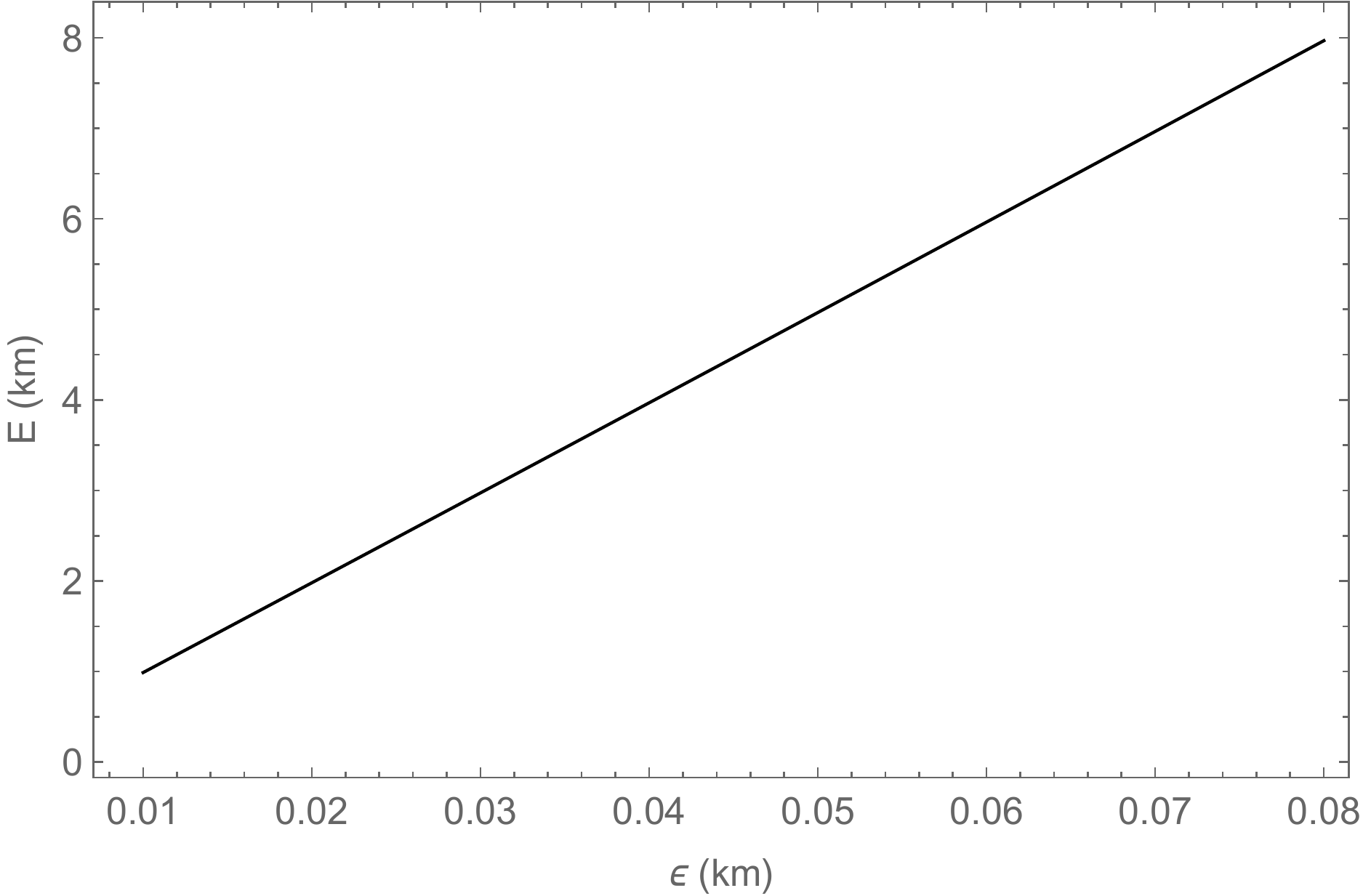}
	\caption{\label{fig4}Variation of Energy $E (km)$ with Shell thickness $\epsilon(km)$ for $a=0.0006$.}
\end{figure}

Fig.~\ref{fig4} shows the increment of energy within the shell with increasing thickness.  
\subsection{Entropy within the Shell}\label{entropy} The stability analysis of an ultra-relativistic comapact object, like Gravastar is very efficiently done by the entropy considerations. The shell region stability is evaluated on the basis of the entropy function of the form gievn below:
\begin{equation}
	S=4\pi\int_{r_1}^{r_2} s(r)r^2e^{\alpha} dr, \label{eq44}
\end{equation}
Here $s(r)=\frac{ak_B}{\hbar}\sqrt{\frac{p}{2\pi}}$ is the entropy density, and from Eq.~\eqref{eq23} we get the total entropy as, 
\begin{eqnarray}
	S&=&\frac{4\pi k_B r_0}{\hbar\sqrt{2\pi}}\int_{r_1}^{r_2} ar^2\sqrt{\frac{a^2}{8\pi}e^{-2ar}}e^{(2+ar)}  dr, \label{eq45} \\
	S&=&4.25\times10^{35}\frac{4\pi k_B r_0}{\hbar\sqrt{2\pi}}a[r^3e^{2+ar}\sqrt{a^2e^{-2ar}}]^{r_2}_{r_1}. \label{eq46}
\end{eqnarray}

\begin{figure}[tbp]
	\centering 
	\includegraphics[width=.45\textwidth]{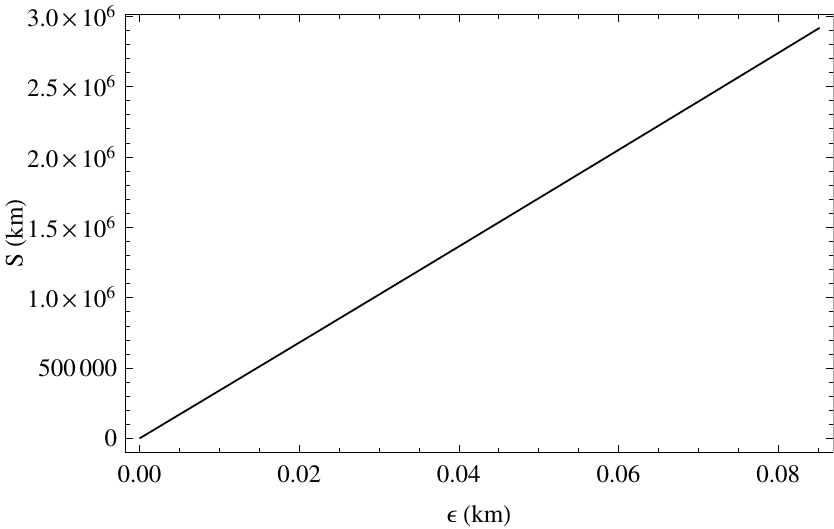}
	\caption{\label{fig5}Variation of Entropy $S (km)$ with Shell thickness $\epsilon(km)$ for $a=0.0006$.}
\end{figure}
From Fig.~\ref{fig5}, we find that the entropy of the gravastar shell approaches zero as the thickness tends to zero. This is considered as a viable condition of stability for a stellar body with single condensate phase \cite{Mazur1}. 
\section{Equation of State (EoS)} \label{eos} 
\begin{figure}[tbp]
	\centering 
	\includegraphics[width=.45\textwidth]{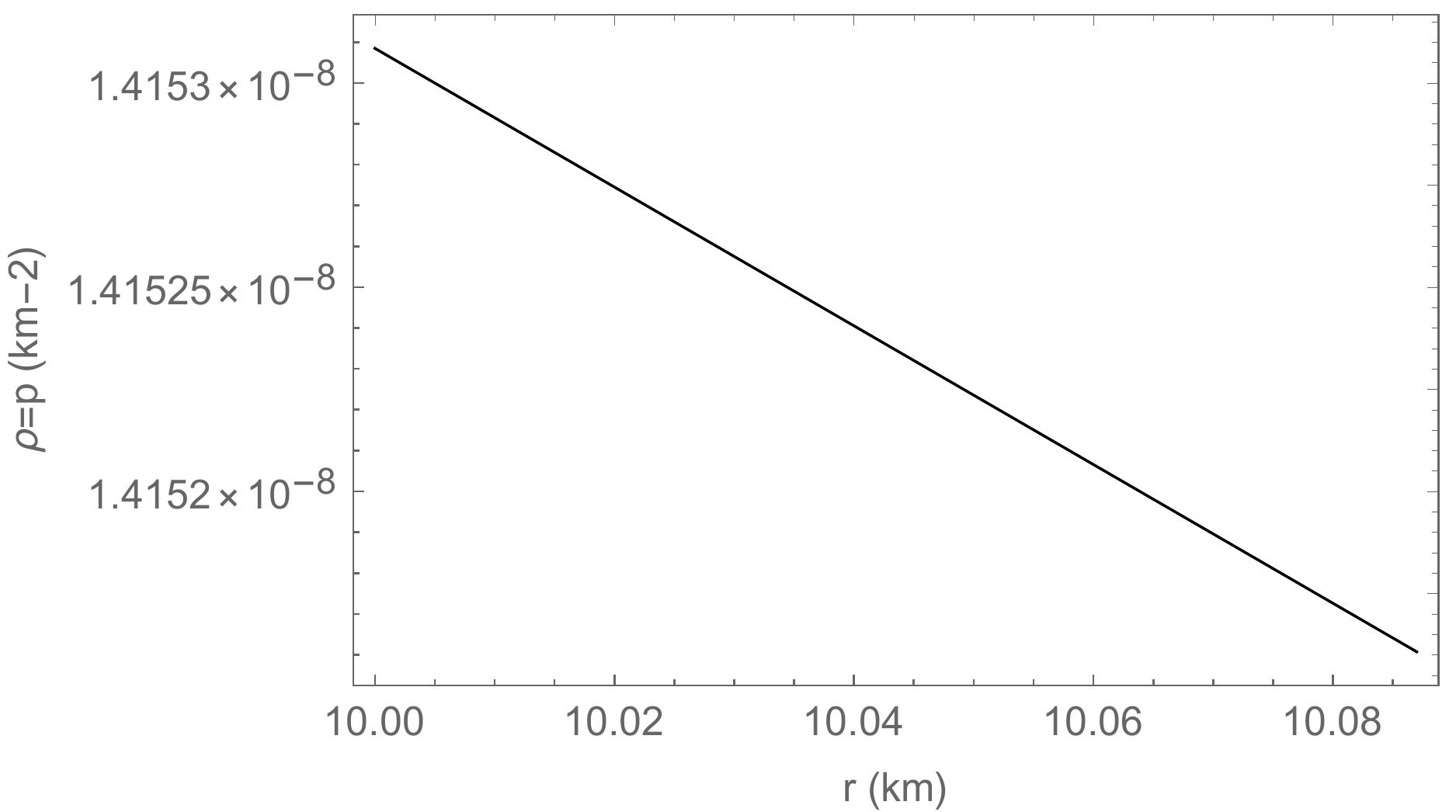}
	\caption{\label{fig8}Variation of EoS with Shell Thickness $\epsilon(km)$.}
\end{figure}

The pressure and density of matter in the thin shell consideration is given by Eq.~\eqref{eq23}. The variation of $p=\rho$ with the increasing shell thickness shows a finitesimally small decreasing pressure and matter density when we proceed from the interior to the exterior region of the gravastar as shown in Fig.~\ref{fig8}.

\section{Conclusion}\label{C} In the paper we investigate gravastars in cylindrically symmetric line element \cite{Bronnikov} and solved the Einstein's field equations \eqref{eq1}. We have also investigated the structure of Gravastars and found that they are in well accordance with the work of Mazur and Mottola \cite{Mazur}. The interior region is free from any singularity ($r=0$) also while the exterior region is the flat, vacuum space-time. The flat exterior approximation is achieved by evaluating the value of Kretschmann scalar $(R_k)$. The two regions are separated by the thin shell of very small but finite thickness with small energy, proving that there is no existence of event horizon. We have also evaluated the mass of the de-Sitter space (Region-I) considering the energy density as the critical density of the universe $\rho_{c}\approx10^{-29} gm/cm^3$ and it is noted that we can neglect the mass of region-I in comparison to the mass of the thin shell (Region-II) for a characteristic choice of radius. Therefore, the shell is responsible for the mass and entropy of the gravastar \cite{Mazur}. We have determined the total mass contained within the thin shell of the gravastar from the parametric choice of constant $C_4$ considered here. In view of Eq.~\eqref{eq26}, it is noted that $C_4$ has some effects on the value of mass of gravastar. Strickingly, with the determination of the mass of the gravastar, we have found that the viable condition for a black hole $(R<2M)$ is obeyed here although it is a singularity free model unlike a black hole. This feature of the model is acceptable for a physically viable gravastar. We have evaluated the proper length of the shell, energy contained within the shell in the context of our model and found that these properties are in well accordance with Mazur-Mottola model \cite{Mazur}. In the Fig.~\ref{fig5}, we have also shown that the entropy of the thin shell approaches a zero value as the thickness of the shell goes to zero. This is another viable condition in favour of stable gravastar modeling, also described by Mazur and Mottola \cite{Mazur1}. With all these features and a consideration of the thin shell as the critical surface of quantum phase transition, gravastars may be considered as a prime alternative manifestation to black holes.

\acknowledgments
DB is thankful to the Department of Physics, Cooch Behar Panchanan Barma University, for the providing the necessary help to carry out the research work. 



\end{document}